\documentclass[%
 reprint,
 amsmath,amssymb,
 aps,
]{revtex4-1}
\usepackage{csquotes}
\usepackage{graphicx}
\usepackage{dcolumn}
\usepackage{bm}
\usepackage{float}
\usepackage{hyperref}
\hypersetup{colorlinks=true, urlcolor=blue, citecolor=blue}

\begin{document}


\title{ Topologically Invariant Double  Dirac States in Bismuth based Perovskites:  Consequence of Ambivalent Charge States and Covalent Bonding }

\author{Bramhachari Khamari, Ravi Kashikar and B. R. K. Nanda}
\email{nandab@iitm.ac.in}
 \affiliation{Condensed Matter Theory and Computational Lab, Department of Physics, Indian Institute of Technology Madras, Chennai, India, 600036}

\date{\today}

\begin{abstract}
Density functional calculations and model tight-binding Hamiltonian studies are carried out to examine the bulk and surface electronic structure of the largely unexplored perovskite family of {\it A}BiO$_3$, where {\it A} is a group I-II element. From the study, we reveal the existence of two TI states, one in  valence band (V-TI) and the other in conduction band (C-TI), as the universal feature of {\it A}BiO$_3$.  The V-TI and C-TI are, respectively, born out of bonding and antibonding states caused by Bi-$\{$s,p$\}$ - O-$\{$p$\}$ coordinated covalent interactions.  Further, we outline a classification scheme in this family where one class follows spin orbit coupling and the other follows the second neighbor Bi-Bi hybridization to induce s-p band inversion for the realization of C-TI states. Below a certain critical thickness of the film, which varies with {\it A}, TI states of top and bottom surfaces couple to destroy the Dirac type linear dispersion and consequently to open narrow surface energy gaps.\\

\end{abstract}

\maketitle

\section{\label{sec:intro}Introduction}

Topological insulators, which are insulating in bulk but with invariant conducting surface states in films \cite{Hsieh,Mele,Hasan,Wray,Pal,Sato,Kubler,Moore,Hughes,Jun,Cheng,Fu,Yan}, have gained considerable attention in the last  decade. As it is a band phenomena arising out of the interplay between crystal and orbital symmetries, there have been extensive studies on the band structures of  a large number of prototypes which include
 Bi based selenides and tellurides \cite{Cheng,Park}, open structures like skutterudites \cite{Ming}, Heusler compounds \cite{Jun,Kubler,Wanx}, and Bi based perovskites ({\it A}BiO$_3$) - {\it A} being an alkali or alkaline earth metal\cite{Clau,Thomale}. The family of perovskites are very distinct from the rest. Firstly, here the topologically invariant (TI) surface state appears when Bi forms the octahedral complex with oxygen. Therefore, perovskites like BiFeO$_3$ with FeO$_6$ complexes are ruled out while the low temperature superconducting materials like KBiO$_3$ and BaBiO$_3$ are actively investigated\cite{Clau,Thomale}. Secondly, unlike the other topological insulators\cite{Ochi,Linyang,Wu}, in {\it A}BiO$_3$ the TI state does not appear at the Fermi level (E$_F$) and instead, it appears far away from the Fermi level (E$_F$) in the conduction band \cite{Clau}. 

There are a number of significant issues that remain unanswered on the formation of TI states in the family of Bi based perovskites. The cause of formation of a TI state in the conduction band\cite{Clau,Thomale}, rather than on the E$_F$, has not been explained. While most of the investigations are restricted to KBiO$_3$ and BaBiO$_3$\cite{Clau,Thomale}, there are many other alkali and alkaline (Sr) elements which are expected to form the perovskite crystal structure of {\it A}BiO$_3$\cite{Kumada, Kazakov}.  Hence a thorough investigation of these compounds will shed light on the underlying physics of the formation of TI states in this family. While in the context of  BaBiO$_3$  a single TI state in the conduction band  has been reported \cite{Clau}, a recent study suggests that in  KBiO$_3$\cite{Thomale} there are indeed two TI states, approximately 10 eV apart with one in the conduction band  and the other in the valence band . It has not been understood and substantiated whether the formation of \enquote{two TI states}
 is a characteristic feature of this family.

In the Bi based TI compounds like tellurides and selenides \cite{Hsieh,Cheng,Park,Hor} the band inversion,  which is a necessary criteria to form TI surface states, has been found to be an outcome of strong spin-orbit coupling (SOC) of the Bi-p states. The band inversion can also occur in some other Bi based compounds \cite{Wray,Jun,Wu,Seiji,Ding} by applying external strain.
In the case of perovskites, while in KBiO$_3$ SOC creates the band inversion\cite{Thomale}, in BaBiO$_3$ the band inversion is present even in the absence of SOC \cite{Clau}. A very recent study on Bi based double perovskites {\it A}$_2$BiXO$_6$, where {\it A} is a divalent cation like Ca, Sr, and Ba, and X is either Br and I, also suggests that the band inversion is not led by SOC \cite{Pi}. Since BiO$_6$ octahedra is a common feature in the crystal structure of these single and double perovskites, it is imperative to devise the mechanism of band inversion in the family of perovskites and classify the systems accordingly. The surface electronic structures so far have been examined using minimal basis set based tight binding (TB) Hamiltonian instead of  a full basis set based first principles calculations \cite{Clau,Thomale}. In such studies, while the surface confinement effects are well taken into account, the microscopic changes in the chemical bonding, which affect the surface states significantly, are often ignored. 
\begin{figure}
\begin{center}
\hspace{-0.6cm}
\includegraphics[angle=-0.0,origin=c,height=15cm,width=6cm]{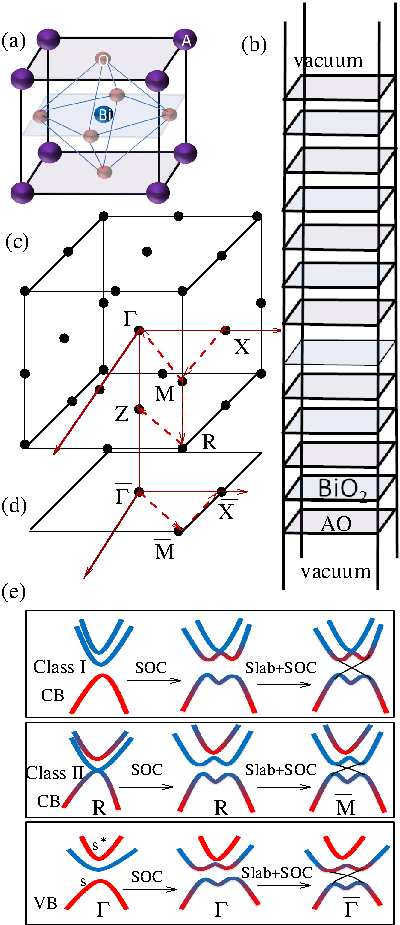}
\caption{(a) Crystal structure of cubic perovskite {\it A}BiO$_3$ with the illustration of BiO$_6$ octahedra. The crystal can also be considered as consisting of {\it A}O and BiO$_2$ planes stacked alternately along the 001 direction. (b) The {\it A}BiO$_3$ slab grown along 001 to calculate the surface electronic states.  The bulk and surface Brillouin zone with high symmetry points are shown in (c) and (d) respectively. (e) Schematic illustration of mechanism of band inversion and formation of topologically invariant surface states. Here, CB and VB stands for conduction and valence band respectively. The class I mechanism of band inversion occurs for {\it A} = Na, K, Rb, Cs and Mg. The rest of the members follow class II mechanism.} 
\label{fig:1}
\end{center}
\end{figure}
This paper examines the bulk and surface band structures of cubic perovskites {\it A}BiO$_3$, where {\it A} is either a monovalent cation (Na, K, Rb, Cs) or a divalent cation (Mg, Ca, Sr, and Ba), to identify the characteristics and the mechanisms that lead to formation of TI states in this family in particular and oxides in general.  The band structures are obtained from  both density functional theory (DFT) and TB calculations. The former is achieved within the framework of full potential and plane wave basis set based FP-LAPW method \cite{Hamann}. The TB Hamiltonian is formulated based on linear combination of atomic orbitals (LCAO) method as developed by Slater and Koster \cite{Koster}. Figure ~\ref{fig:1} summarizes the important findings in this work. Two TI states, one in the  valence band and other in the conduction band  are characteristic features of Bi based perovskites. While the TI state in the valence band (TI-V) is formed by the bonding states, resulted from O-$\{p\}$ - Bi-$\{s,p\}$ strong covalent hybridization, the TI state in the conduction band, (TI-C) is formed by the corresponding antibonding states. The band inversion for TI-V occurs through SOC of Bi-p states. However, the band inversion of the TI-C state is either created by the SOC, as in the case for Na, K, Rb, Cs or Mg based perovskites, or through weak but sensitive second neighbor Bi-p - Bi-p interactions as in the case for Ca, Sr and Ba based perovskites.

\section{\label{sec:method}Structural and Computational Details
$\mathbf{{\it A}BiO_3}$ }

The experimentally synthesized {\it A}BiO$_3$ members either crystallize in the cubic phase\cite{Sle} or in the slightly distorted cubic (monoclinic) phase\cite{Cox}. However, since the TI states are not expected to be affected by minor distortion in the lattice\cite{Clau}, we have carried out the calculations in the cubic phase (space group {\it Pm-3m}) as shown in Fig.~\ref{fig:1}(a).  The salient features of the cubic structure are: (i) BiO$_6$ forms a perfect octahedral complex and (ii) the crystal structure has alternately stacked {\it A}O and BiO$_2$ planes. To study the surface states, slabs along 001 are constructed as illustrated in Fig.~\ref{fig:1}(b). Both  top and bottom surfaces are terminated with {\it A}O planes. The band structures are examined as a function of slab thickness to examine the evolution of the TI states and possible interactions between them. Unlike BaBiO$_3$\cite{Sle}, KBiO$_3$\cite{Sinha,Sahra} and  RbBiO$_3$\cite{Tomeno}, for the rest of the family, the cubic experimental phase is not well established. Therefore, the lattice parameter of BaBiO$_3$ is used to calculate the band structure of other members. Such an assumption is not expected to affect the qualitative conclusions made in this paper. 

The density functional calculations are carried out using the full potential linearized augmented plane wave (FP-LAPW) formalism \cite{Hamann} as implemented in the WIEN2k simulation package \cite{Aug}.
Augmented plane waves in the interstitial and localized orbitals within the muffin-tin sphere are used to construct the basis sets. The largest vector in the plane wave expansion is obtained by setting R$K_{{\it max}}$ to  7.0. The PBE-GGA exchange-correlation functional is used to solve the Kohn-Sham equations \cite{Perdew}. To carry out the Brillouin-zone integration $10\times{10}\times{10}$ k mesh yielding 35 irreducible k points, is used for the bulk structure and a proportionate k mesh is used for the slab calculation. The details of the TB Hamiltonian of Eq. (1) are presented in the appendix.

\section{\label{sec:method}Bulk Electronic Structure of $\mathbf{{\it A}BiO_3}$ }

\begin{figure}
\begin{center}
\hspace{-0.2cm}
\includegraphics[angle=-0.0,origin=c,height=8cm,width=8cm]{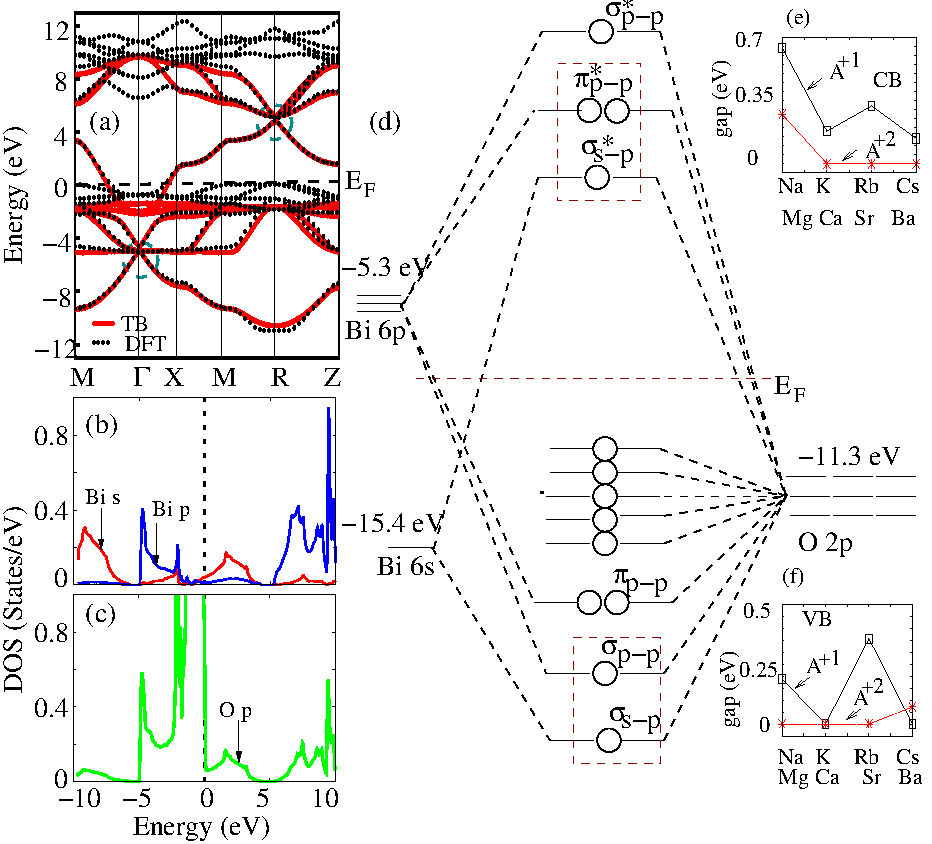}
\caption{Bulk electronic structure of KBiO$_3$, in the absence of SOC, illustrated through (a) band structure, (b) and (c) partial DOS, and (d) schematic molecular orbital picture (MOP). The band structure is obtained from both DFT (black dashed lines) and exact diagonalization of the TB Hamiltonian of Eq. (1) (red solid lines). In MOP, $\{\sigma, \pi\}$ and $\{\sigma^*, \pi^* \}$, respectively, represent the bonding and antibonding states which are resulted from the nearest neighbor Bi-$\{s, p\}$ - O-$p$ covalent hybridization. Each member of {\it A}BiO$_3$ has similar electronic structure except the fact that there is an upward shift in E$_F$ as the charge state of {\it A} changes from +1 to 
+2.   The  antibonding states $\sigma^*_{s-p}$ and $\pi^*_{p-p}$ either touch each other to create accidental degeneracy or open a gap between them depending on {\it A} as shown in (e). Similarly, the bonding states $\sigma_{s-p}$ and $\sigma_{p-p}$ either touch each other or open a gap between them depending on {\it A} as shown in (f). }
\label{fig:2}
\end{center}
\end{figure}

The chemical bonding plays a significant role in deciding the transport properties of a given solid. In most of the reports on  Bi based alloy topological insulator\cite{Hai}, Bi based tellurides and selenides \cite{Hsieh,Pal,Park,Hor}, the nature of chemical bonding has not been emphasized adequately while explaining the formation of TI states.  However, the nature of chemical bonding in Bi based perovskites cannot be overlooked. Firstly because the TI state does not occur at E$_F$ and hence its formation needs to be understood. Secondly, the electronic structure of perovskites exhibits highly dispersive bands that are very different from that of the Bi alloys \cite{Haijun,Luo}. We will first present the bulk electronic structure of KBiO$_3$ as a prototype and extend the understanding to the other members of the family. 
    
  Bulk electronic structure of KBiO$_3$ can be best explained from Fig.~\ref{fig:2}, where the  band structure, partial densities of states (DOS), and the resulted (schematic) molecular orbital picture are shown. The DFT band structure [Fig.~\ref{fig:2}(a), black dotted lines] suggests that this metallic system has two sets of  four highly dispersive states - one lies in the conduction band and the other lies in the valence band spectrum. The highly dispersive states of the conduction band nearly touch each other at the high symmetry point R as marked with a dotted circle. The orbital projected DOS infers that the lower lying band is primarily of Bi-s character and the remaining three bands are primarily of Bi-p character.  
  Similarly, in the valence band, the dispersive states nearly touch each other at $\Gamma$ with the lower band being Bi-s and upper bands being Bi-p in nature. Though a similar feature has been reported in earlier studies \cite{Thomale}, the separation between the dispersive states of the conduction band and valence band which can be as large as 10 eV has not been analyzed and understood. Such exceptionally large separation, which is not observed in other Bi based TI families \cite{Cheng,Park} can only happen provided there is a strong covalent interaction making a set of bonding dispersive states lying lower in energy and a set of antibonding dispersive states lying higher in energy.  Since, intrasite orbital overlap cannot exist, the  interaction between Bi-s and Bi-p states is ruled out. On the contrary, due to the symmetric BiO$_6$ octahedra, stronger  interaction is expected between the Bi-$\{$s,p$\}$ and O-$\{$p$\}$ states. Indeed the partial DOS, plotted in Fig.~\ref{fig:2}(c), shows significant presence of O-p characters everywhere in the energy spectrum. 
  
In order to quantify the covalent interaction between the Bi-$\{$s, p$\}$ and O-p states  we have constructed a tight-binding model Hamiltonian with minimal basis set and is based on LCAO as adopted by Slater and Koster \cite{Koster}.  
  
  The Hamiltonian, which also includes SOC, is expressed as:
  \begin{equation}
H = \sum_{i,\alpha}\epsilon_{i\alpha}c_{i\alpha}^\dag c_{i\alpha} + \sum_{ij;\alpha,\beta}t_{i\alpha j\beta}(c_{i\alpha}^\dag c_{j\beta} + h.c) + \lambda\textbf{L}\cdot\textbf{S}
\end{equation}
  
  Here {\it i} and {\it j} are the site indices while $\alpha$ and $\beta$ are the orbitals (Bi-$\{$s, p$\}$ and O-p) forming the basis set. The parameters $\epsilon_{i\alpha}$ and $t_{i\alpha j\beta}$, respectively, represent the on-site energy and hopping integrals. In addition to the nearest neighbor Bi-O interactions, the second neighbor Bi-Bi interactions are included in the Hamiltonian.
   The third term of the Hamiltonian represents spin-orbit coupling (SOC) among the Bi-p states with coupling strength $\lambda$. Further details of the TB model, e.g.,  Hamiltonian matrix, optimized values of $t_{i\alpha j\beta}$, and SOC strength $\lambda$, are presented in the Table II of the appendix. 
   
  The Hamiltonian is diagonalized and the resulted bands are fitted with that of DFT. The TB bands in the absence of SOC are plotted in red solid lines in Fig.~\ref{fig:2}(a). We find excellent agreement  between the TB bands and the two sets of highly dispersive (HD) DFT bands. In fact at $\Gamma$ and R, the DFT and TB bands coincide. For KBiO$_3$ the values of nearest neighbor (NN) hopping interaction strengths V$^{BiO}_{sp\sigma}$, V$^{BiO}_{pp\sigma}$, and V$^{BiO}_{pp\pi}$ are found to be 2.1, 2.95 and -0.75 eV respectively. The value of next nearest neighbor (NNN) hopping interaction strengths V$^{BiBi}_{ss\sigma}$, V$^{BiBi}_{sp\sigma}$, V$^{BiBi}_{pp\sigma}$ and V$^{BiBi}_{pp\pi}$ are found to be -0.04, 0.02, -0.08, and 0.16 eV, respectively. Similar results are obtained for other members and details are listed in Table I of the appendix. The electronic structure and chemical bonding in the family of {\it A}BiO$_3$ as inferred from the DFT and TB calculations are summarized in the molecular orbital picture (MOP) shown in Fig.~\ref{fig:2}(d). The cation {\it A} transfers its valence electrons to the anion O and does not take part in the bonding. However, Bi remains in an ambiguous charge state since electrons from the outer Bi-s orbitals, with -15.4 eV atomic on-site energy,  cannot be transferred to O-p states which are at higher on-site energy of -11.3 eV. Therefore, electron sharing occurs between Bi and O through strong covalent interaction. The Bi-s - O-p interaction creates a bonding state $\sigma_{s-p}$ and an antibonding state $\sigma^{*}_{s-p}$. Similarly, the Bi-p - O-p interactions create a set of bonding states $\{ \sigma_{p-p}, \pi_{p-p}\}$ and a set of antibonding states $\{ \sigma^{*}_{p-p}, \pi^{*}_{p-p}\}$ as shown in Fig.~\ref{fig:2}(d). The bonding states $\sigma_{s-p}$ and $\sigma_{p-p}$ either touch each other, to create accidental degeneracy, or open a gap between them depending on the cation {\it A} (see Fig.~\ref{fig:2}(f)). Similarly the  antibonding states $\sigma^*_{s-p}$ and $\pi^*_{p-p}$ either touch each other or open a gap between them depending on {\it A} as shown in Fig.~\ref{fig:2}(e). As a consequence, there is a possibility of two TI states, one in the valence band (VB) and the other in the conduction band (CB), in the family of {\it A}BiO$_3$. 
  
  The MOP is nearly similar for each member of {\it A}BiO$_3$. To avoid the repetition, we have shown the band structure of these compounds (other than KBiO$_3$) in Fig. 7 of the appendix. However, in the coming sections we will see that even though the NNN Bi-Bi  hopping interaction strengths are weak, a minor variation of them can lead to different mechanisms for forming the TI states in the conduction band. 

\begin{center}
\begin{figure*}
\includegraphics[angle=-0.0,origin=c,height=12cm,width=16cm]{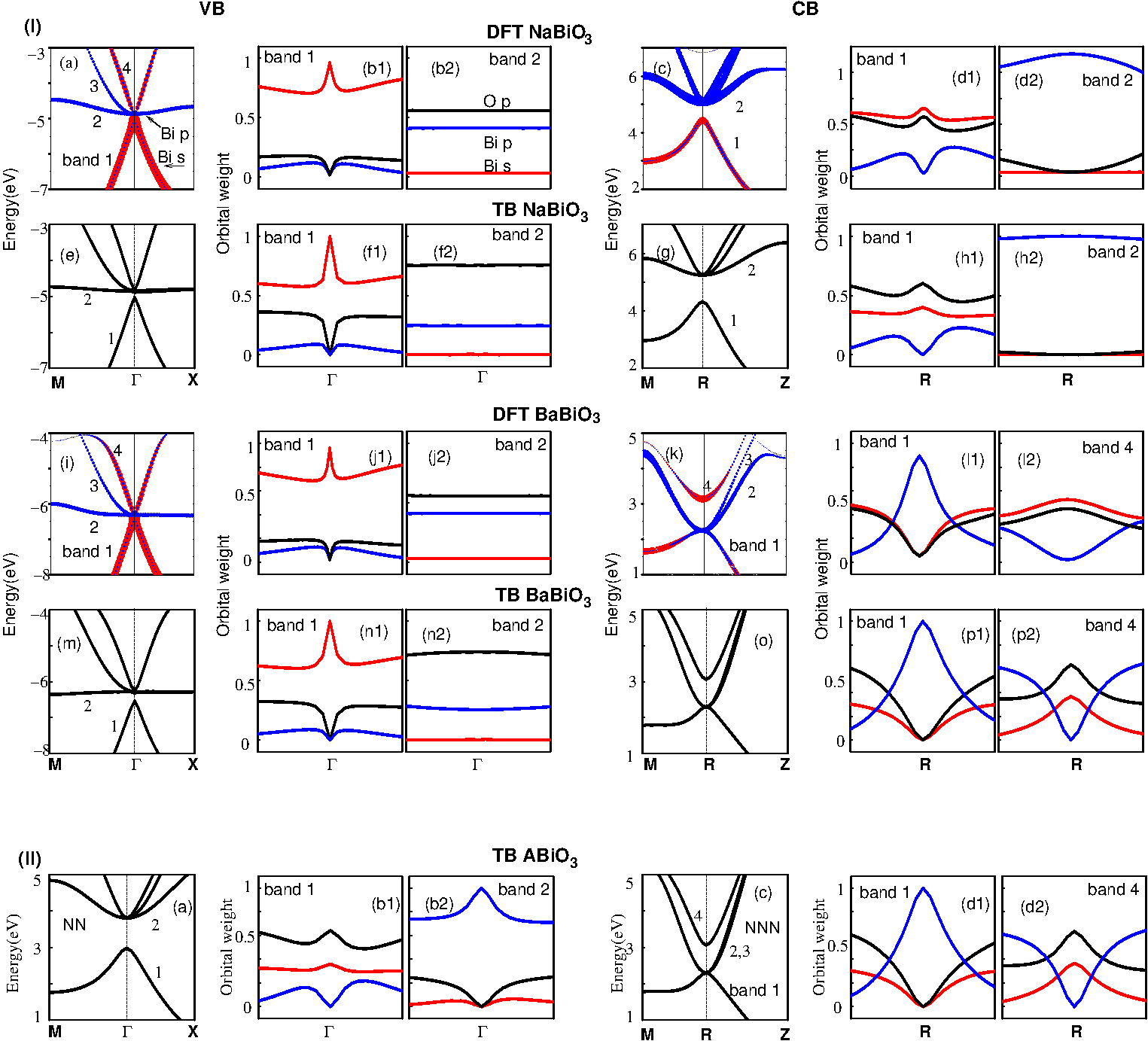}
\caption{(I) DFT and TB obtained valence bands (VB) along M-$\Gamma$-X (first column) and conduction bands (CB) along M-R-Z (fourth column) for NaBiO$_3$ and BaBiO$_3$ in the absence of SOC. The DFT bands are shown with projected Bi-s and p orbital characters. Right to each band structure the orbital weights (contribution of individual orbitals in forming a band) of the relevant bands, responsible for constructing the topologically invariant surface states, are shown. For the VB, the relevant ones are band-1 and band-2. For the CB, the relevant ones are band-1 and band-2 for NaBiO$_3$ and band-1 and band-4 for BaBiO$_3$. The orbital weights of band-1 and band-2 for VB are shown in column 2 and 3, respectively.  Similarly  the orbital weights of band-1 and band-2 or band-4 of CB are shown in column 5 and 6, respectively. 
 (II) TB obtained BaBiO$_3$ conduction band structure: (a)when the interactions are restricted to nearest neighbor (NN) and (c) when they include Bi-Bi next nearest neighbor (NNN) interactions. The orbital weights relevant to (a) are shown in  (b1) and (b2) and relevant to (c) are shown in (d1) and (d2). The panel (II) infers that  reasonable NNN Bi-Bi interactions can induce band inversion in the CB spectrum. However, later we will find that with inclusion of SOC, the band inversion will occur even with negligible NNN interactions. The results are extendable to each member of {\it A}BiO$_3$. }
\label{fig:3}
\end{figure*}
\end{center} 
\subsection{Hybridization Induced Band Inversion and Classification of ABiO$_3$ compounds}
The broad picture of electronic structure of {\it A}BiO$_3$ presented in Fig.~\ref{fig:2}  is inadequate to evolve a mechanism for the formation of TI surface states in this family. We need to investigate the local variation in the valence band dispersion at the high symmetry point $\Gamma$ and conduction band dispersion at $R$  and also identify the orbitals constructing these bands. Therefore, in Fig.~\ref{fig:3} we have plotted the DFT and TB obtained conduction bands along M-R-Z and valence bands along M-$\Gamma$-X for NaBiO$_3$ and BaBiO$_3$. These two are chosen as prototypes to represent the compounds with {\it A}$^{+1}$ and {\it A}$^{+2}$ cations.  

{\it Valence band structure} -- The orbital projected band structure (orbital contribution to a given band is proportional to thickness of the curve shown), obtained from DFT, shows that the valence bands in the vicinity of $\Gamma$ for both the compounds  are  nearly exact with the lower lying band (band-1) is predominantly of Bi-s and the upper two weakly disperse bands (band-2, band-3) are of Bi-p character [see Figs. 3(a) and (i)]. The uppermost highly dispersive band (band-4) is occupied by Bi-s orbital in the vicinity of $\Gamma$. However, in the rest of the Brillouin zone (not shown here) it is more dominated by Bi-p characters. 

{\it Conduction band structure} -- From the analysis of the band dispersion and orbital weight factor, we find that in the case of conduction bands, there is a clear distinction between NaBiO$_3$ [Figs. ~\ref{fig:3}I (c,d1,d2 DFT) and (g,h1,h2 TB)] and  BaBiO$_3$ [Figs. ~\ref{fig:3}I(k,l1,l2 DFT) and (o,p1,p2 TB)]. For the former, at   $R$, the lower band, is  occupied by Bi-s, while the upper three bands are occupied by Bi-p characters. On the contrary for the latter, the lower band, is of Bi-p character while the uppermost band (band-4) is of Bi-s character.
Therefore,  it suggests that there is a hybridized induced band inversion in BaBiO$_3$ between the lower band and the uppermost band. In addition to BaBiO$_3$, the orbital weights listed in Table III of the appendix, suggests similar band structure in the case of CaBiO$_3$ and SrBiO$_3$. Interestingly a very recent article \cite{Pi} suggests that   {\it A}$_2$BiXO$_6$ (X = Br, I) exhibit s-p band inversion without SOC when {\it A} is either Ca, Sr, or Ba. This carries significance as in most of the Bi based topological insulators, the band inversion  occurs only through spin orbit coupling. 

To provide a quantitative picture of hybridized band inversion,  both our  DFT [Fig.~\ref{fig:3}I(l1,l2)] and TB [Fig.~\ref{fig:3}I(p1,p2)] calculations show that  away from $R$, the band-1 (lower band) is composed of 40$\%$ Bi-s and 60$\%$ O-p character. But at $R$, the band is completely formed by Bi-p orbitals. In the case of band-4 (uppermost band) it is  reversed. Away from $R$, the band is composed of 60$\%$ Bi-p and 40$\%$ O-p states. At $R$ it is composed of 40$\%$ Bi-s and 60$\%$ O-p. This establishes the Bi s-p band inversion at $R$ in BaBiO$_3$. The valence bands of both the compounds appear similar and no band inversion is observed between band-1 and band-2 as in the case of conduction bands of NaBiO$_3$. 
The calculation of orbital weights also shows that there is a significant presence of O-p characters in the conduction and valence dispersive bands which in turn implies that in  perovskites, unlike the selenides and tellurides, the hybridized states, instead of pure Bi characters, construct the TI states.

The cause of hybridized induced band inversion in BaBiO$_3$ and the  absence of it in NaBiO$_3$ can be explained by solving the  tight-binding Hamiltonian without SOC. In Fig.~\ref{fig:3}II(a-d2), we have plotted the conduction band structure of {\it A}BiO$_3$, around the high symmetry point $R$, and the orbital weight factors in these bands for two cases. In the first case, only NN Bi-O interactions are included in the TB Hamiltonian. The resulting band structure as well as the plot of k-dependent orbital weights [Figs. ~\ref{fig:3}II(a) and (b1,b2)]  resembles that of NaBiO$_3$ [see Figs.~\ref{fig:3}I(g) and (h1,h2)]. However, when the NNN Bi-Bi interactions are included in the Hamiltonian, the resulted band structure as well as the plot of the orbital weights [Figs.~\ref{fig:3}II(c) and (d1,d2)] resembles that of the BaBiO$_3$ [see Figs.~\ref{fig:3}I(o) and (p1,p2)]. This suggests that the second neighbor interactions induce the s-p band inversion. From the optimized TB parameters listed in Table I of the appendix, we find that the second neighbor interactions, particularly the pp$\sigma$ and pp$\pi$, are significant in CaBiO$_3$, SrBiO$_3$, and BaBiO$_3$. Accordingly, in the absence of SOC, the s-p band inversion occurs in these compounds while the rest of the compounds do not have it. We may note that in general, the critical second neighbor Bi-Bi interaction strengths (t$_{ss}$, t$_{sp}$, t$_{pp\sigma}$, and t$_{pp\pi}$), above which the hybridization induced inversion occurs, vary with the cation {\it A}.   

\subsection{Effect of spin-orbit coupling}
\begin{center}
\begin{figure*}
\includegraphics[angle=-0.0,origin=c,height=11cm,width=15.0cm]{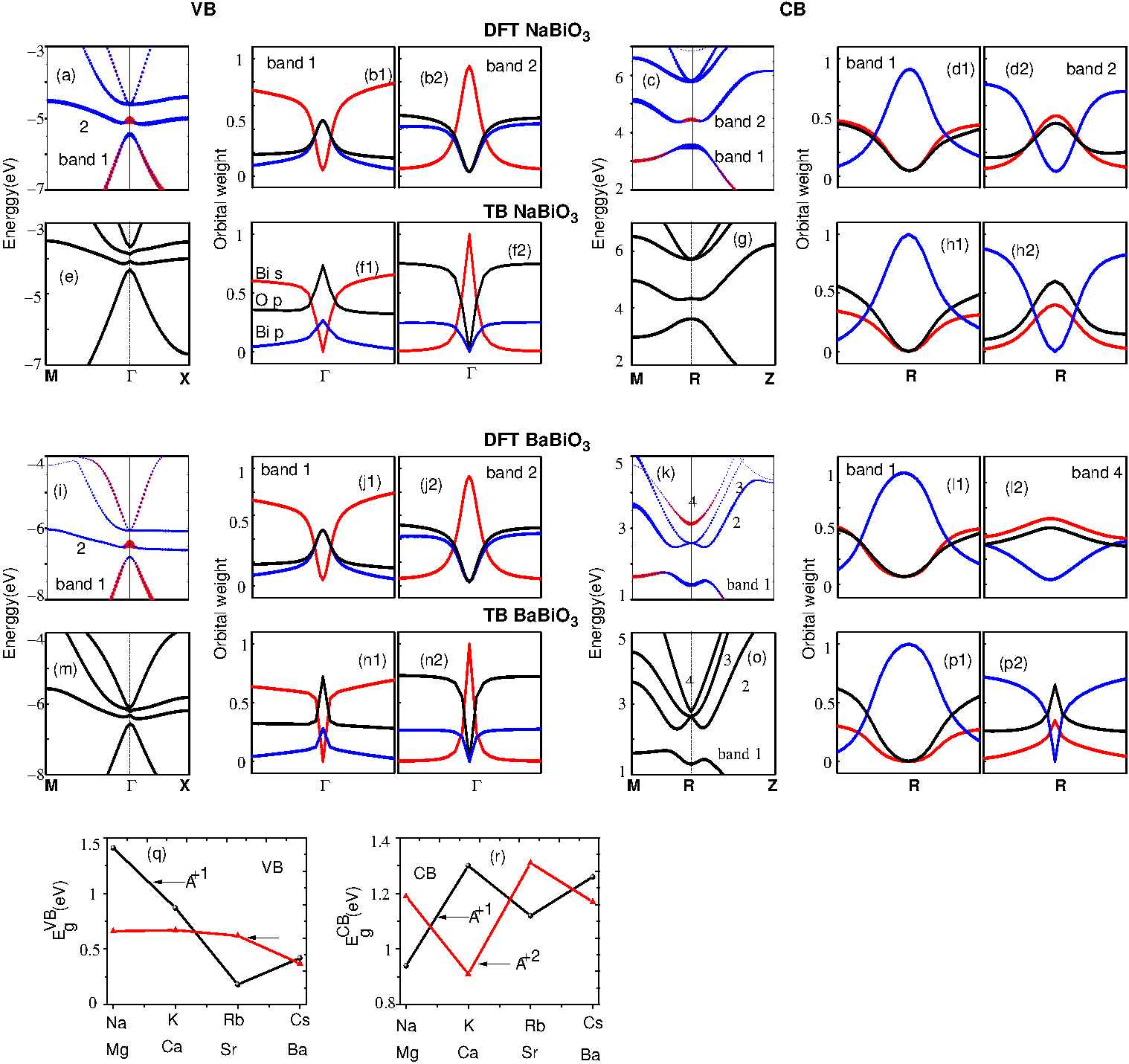}
\caption{
DFT+SOC and TB+SOC obtained valence bands along M-$\Gamma$-X (first column) and conduction bands along M-R-Z (fourth column) for NaBiO$_3$ and BaBiO$_3$. The DFT bands are shown with projected orbital characters. Right to each band structure the orbital weights of the relevant bands are shown.
For the VB, the relevant ones are band-1 and band-2. For the CB, the relevant ones are band-1 and band-2 for NaBiO$_3$ and band-1 and band-4 for BaBiO$_3$. The orbital weights of band-1 and band-2 for VB are shown in column 2 and 3 respectively.  Similarly  the orbital weights of band-1 and band-2 or band-4 of CB are shown in column 5 and 6, respectively.  With inclusion of SOC, the s-p band inversion occurs between the band-1 and band-2 in the valence band of both the compounds and in the conduction band of NaBiO$_3$. The band inversion (between 1 and 4) in the conduction spectrum of BaBiO$_3$ remains unaffected by SOC. The SOC either creates a gap or amplifies the already existing gap at R in the conduction band (E$^{CB}_g$) and at $\Gamma$ in the valence band (E$^{VB}_g$). [(q) and (r) change] plot (E$^{CB}_g$) and (E$^{VB}_g$), respectively, for different {\it A} cations.}
\label{fig:4}
\end{figure*}
\end{center}  

Having understood the electronic structure due to chemical bonding through Figs. 2 and 3,  in this subsection we shall discuss the effect of SOC of the Bi-p states on the band structure. Figure.~\ref{fig:4} shows the DFT + SOC and TB + SOC  band structures as well as k-dependent orbital weights of the relevant bands for NaBiO$_3$ and BaBiO$_3$.  The SOC introduces two significant changes in the band structure at the high symmetry point $\Gamma$ in the valence band and at $R$ in the conduction band. Firstly either a gap appears at these points or the already existing gap [see Figs. ~\ref{fig:2}(e) and (f)]  gets amplified except in the case of RbBiO$_3$. The magnitude of the gap at $\Gamma$ (E$^{VB}_g$) and at R (E$^{CB}_g$) as a function of {\it A} cation is shown in Figs.~\ref{fig:4}[(q) and (r)], respectively. For RbBiO$_3$, SOC reduces the band gap from 0.35 to 0.2 eV. However, the gap remains robust. Secondly, from the calculations of orbital weights, the band-1 and band-2 of valence band alter their characters at $\Gamma$. For example, according to our TB calculations on NaBiO$_3$, without SOC [see Fig.~\ref{fig:3}I(f1,f2)], at $\Gamma$ band-1 is composed only of Bi-s character and band-2 is composed of 25$\%$ Bi-p and 75$\%$ O-p characters. With the inclusion of SOC [see Fig.~\ref{fig:4}(f1,f2)], the band-1 is now composed of 27$\%$ Bi-p and 73$\%$ O-p characters, whereas band-2 is solely made up of Bi-s states. The DFT calculations provide similar results as can be seen from Figs.~\ref{fig:3}I(b1,b2) and ~\ref{fig:4}(b1,b2). Band-1 and band-2 of the conduction band (for {\it A} = Na, K, Rb, Cs, and Mg) also undergo similar inversion at R. Table IV of the appendix, lists the  orbital weights at $\Gamma$ and $R$ for band-1 and band-2 to reconfirm the band inversion in these compounds. The hybridization induced band inversion in the conduction band of BaBiO$_3$, CaBiO$_3$, and SrBiO$_3$ remains unaffected by SOC. By fitting the TB+SOC bands with that of the DFT+SOC obtained bands, we found that the SOC strength for the latter three compounds is about 0.5 eV which is nearly 0.2 eV smaller compared to that of the other members of the family.

\section{Surface Electronic Structure of $\mathbf{ABiO_3}$}

The hallmark of topological insulators is the formation of a Dirac type surface state within the bulk band gap created by SOC. Such a state is invariant under adiabatic deformation\cite{Chao}. As bulk ABiO$_3$ exhibit two SOC induced/amplified bulk band gaps (E$^{VB}_g$ and E$^{CB}_g$) it is imperative to see whether this perovskite family forms two such Dirac type surface states protected by symmetry. 

These surface states, so far, are primarily examined by solving the TB model Hamiltonians with minimal basis set instead of a full-basis set based DFT study \cite{Clau,Thomale}. In most of the cases, the minimal Wannier basis is obtained from the bulk electronic structure and subjected to the TB Hamiltonian to obtain the {\it k}-dependent  eigenstates. In such studies while the surface confinement is well taken into account, the microscopic changes in the chemical bonding, which can bring significant changes in the surface states, are often ignored. Therefore, in this section we present the surface electronic structure calculated using the FPLAPW+Lo method where both plane waves and local orbitals form the basis.  

Furthermore, recent reports suggest that not necessarily all the films of well established topological insulators exhibit Dirac-like surface states. From the ARPES study it is found that in ultra thin Bi$_2$Se$_3$ (three quintuple layers and less), a gap appears between the topological surface state bands \cite{Sakamoto}. The first principles calculations infer similar conclusions in the case of Bi$_2$Te$_3$ films \cite{Park}. Keeping this in mind we have examined the robustness of the TI surface state of {\it A}BiO$_3$ as a function of film thickness. 

\begin{widetext}

\begin{figure}[h]
\begin{center}
\hspace{-0.2cm}
\includegraphics[angle=-0.0,origin=c,height=7.0cm,width=16.0cm]{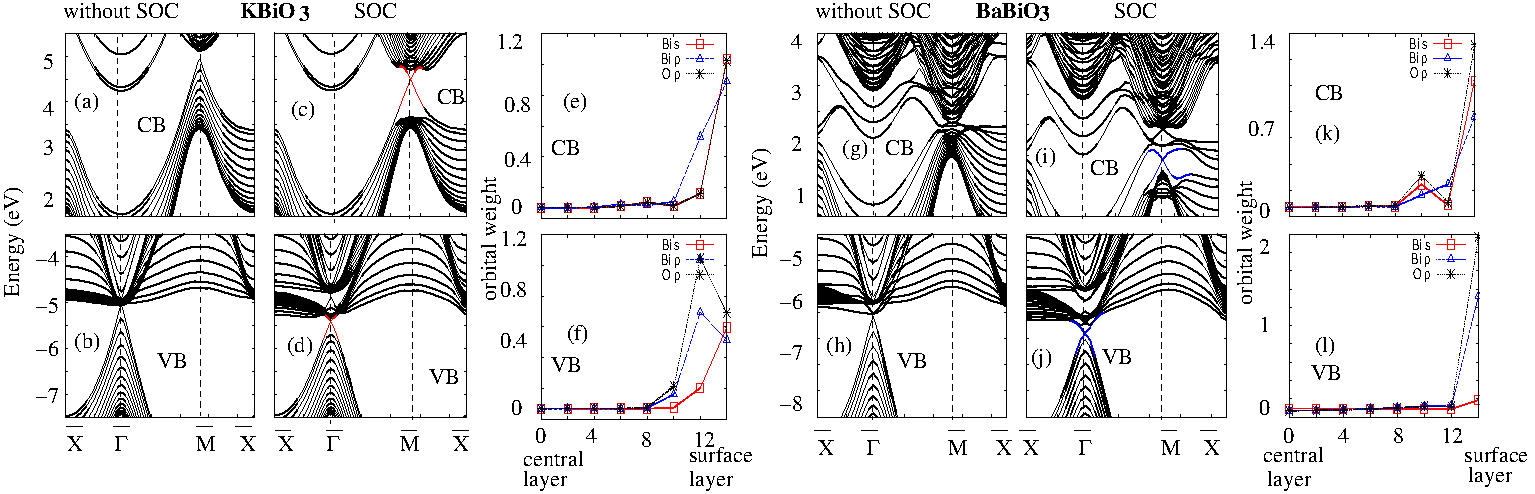}
\caption{ Valence and conduction band structure of 15 unit cell thick KBiO$_3$ [(a) - (d)] and BaBiO$_3$[(g) - (j)] without and with SOC. The layer resolved orbital weights,  in the vicinity of $\bar{M}$  for conduction band and of $\bar{\Gamma}$ for valence band, corresponding to the SOC band structure are also shown in (e), (f) for KBiO$_3$ and in (k), (l) for BaBiO$_3$. Here we gather that orbitals of the first two to three layers from the surface constitute these linear bands.}
\label{fig:5}
\end{center}
\end{figure}

\begin{center}
\begin{figure*}
\hspace{0.5cm}
\includegraphics[angle=-0.0,origin=c,height=7.5cm,width=14.0cm]{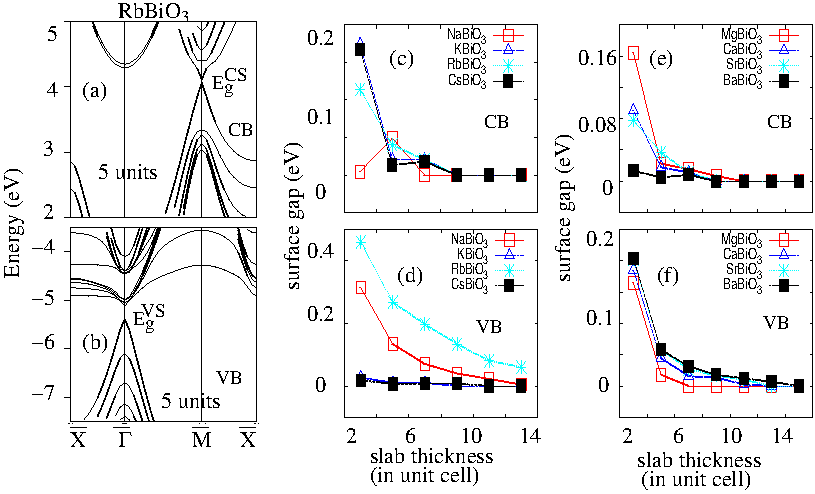}
\caption{Valence (a) and conduction (b) band structure for 5 unit cell thick RbBiO$_3$ demonstrating the existence of a band gap between the topological surface states both in conduction (E$_{g}^{CS}$) and valence band spectrum (E$_{g}^{VS}$). The surface band gap lies between the SOC induced bulk gap (E$_{g}^{VB}$ or E$_{g}^{CB}$). Figures (c) and (d) plot the variation of E$_{g}^{CS}$ and E$_{g}^{VS}$ as a function of film thickness for ABiO$_3$ when A is in +1 charge state (Na, K, Rb. and Cs). Figures (e) and (f) are the same as (c) and (d) but when {\it A} is in +2 charge state (Mg, Ca, Sr, and Ba).}
\label{fig:6}
\end{figure*}
\end{center}

\end{widetext}
First, to examine the effect of chemical bonding, we shall analyze surface electronic structure without SOC. To make the point, the band structure of 15 unit cell thick KBiO$_3$ and BaBiO$_3$ slabs are shown in Fig.~\ref{fig:5}. The figure suggests that while the bulk valence band gap at $\Gamma$ ($\bar{\Gamma}$) and bulk conduction band gap at R (in the surface Brillouin zone R is mapped to $\bar{M}$) are either reduced or disappeared completely in this slab, there is no signature of Dirac type bands appearing at  $\bar{\Gamma}$ and $\bar{M}$ of the surface Brillouin zone. Therefore, the hybridized induced band inversion has a minimal role in forming the TI surfaces states.  With the inclusion of SOC [see Figs. ~\ref{fig:5} (c, d) and (i, j)] we find that linearly dispersed bands, akin to the TI states, appear within the bulk gap. The layer and atom resolved orbital weights of these linear bands estimated at $\bar{M}$ or $\bar{\Gamma}$  [see Figs. ~\ref{fig:5}(e) and (f) for KBiO$_3$; (k) and (l) for BaBiO$_3$] imply that these linear bands are made up of surface O-p, Bi-s, and Bi-p, states. As we move from the surface to the interior, the contribution rapidly vanishes and three layers below the surface, the contribution to these linear bands ceases to exist. Therefore, these linear bands are well defined surface states and satisfy all the criteria to be called TI states. The surface TI states of each member of this family  are shown in Fig. 8 of the appendix.
\\

The robustness of the surface TI states can be measured based on two factors: (i) its invariance with adiabatic deformation \cite{Chao} and (ii) its invariance with respect to the thickness of the film. While the former has been proved based on parity of the bands and calculation of Chern numbers (Z$_2$)\cite{Fu}, the latter is purely a function of chemical bonding which has not been highlighted yet for the family of perovskites. 
\\

The band structure of five unit cell thick RbBiO$_3$, shown in Figs. ~\ref{fig:6}(a) and ~\ref{fig:6}(b) reveals that the linear conduction bands at $\bar{M}$ give rise to a gap (E$_g^{CS}$) of 0.17 eV. Similarly the linear valence bands at $\bar{\Gamma}$ give rise to a gap (E$_g^{VS}$) of 0.45 eV. Both of these gaps lie well within the SOC driven bulk band gaps. Furthermore, from the layer resolved orbital weights (not shown here) we find that the bands below and above  of these new energy gaps are formed by the orbitals of the surface layers. This implies the presence of a strong interaction between the bottom and top surface states which gives rise to these gaps. To provide a quantitative measure, in Fig.~\ref{fig:6}(c)-~\ref{fig:6}(f)we have plotted E$_g^{VS}$ and E$_g^{CS}$ as a function of film thickness. As expected, we find that the gap sharply decreases with increasing thickness. However, the critical thickness below which the gap appears, varies with the cation {\it A}. Also for a given compound the critical thickness differs for formation of E$_g^{VS}$ and E$_g^{CS}$. With increase in the thickness of the slab, the separation between the bottom and top surfaces increases and thereby the coupling between the surface states vanishes to create two noninteracting TI Dirac bands.

\section{Summary and Conclusion}

To summarize, the bulk and surface electronic structure of the {\it A}BiO$_3$ family, where {\it }A is either a monovalent (Na, K, Rb, and Cs) or a divalent (Mg, Ca, Sr, and Ba) element, are obtained from DFT calculations and from exact diagonalization of a minimal basis based TB Hamiltonian. Existence of two topologically invariant (TI) surface Dirac states, one in the conduction band and the other in the valence band, is discovered to be the characteristic feature of this family. The valence TI state arises from the bonding interaction between  Bi-$\{$s,p$\}$ - O-$\{$p$\}$ states in   the BiO$_6$ octahedral complex.  The corresponding antibonding interaction constitutes the conduction TI state. While the usual spin orbit coupling (SOC) induces s-p band inversion in the valence band, variance in it in the case of conduction band classifies the perovskite family into two where one class ({\it A} = Na, K, Rb, Cs and Mg) follows the SOC mechanism and the other ({\it A} = Ca, Sr and Ba) follows the  hybridization induced inversion mechanism for the inversion of Bi-s and p characters in the conduction band. The top and bottom surface TI states of {\it A}BiO$_3$ can couple to destroy their Dirac type linear dispersion. The critical thickness (C$_T$), above which the coupling ceases to exist, varies with {\it A} and lies in the range 4 to 14 unit cell. C$_T$ is also different for conduction and valence surface states. The mechanisms of band inversion and formation of more than one TI surface Dirac states as well as the coupling between these states, proposed in this paper, can be extended to other complex oxides  where the heavy elements like Bi and Pb form symmetric oxygen complexes. 

Shifting the Dirac states from  conduction and valence bands to the Fermi surface, through chemical doping or carrier doping, will make these oxide families viable for futuristic devices. In a very recent study by X. Zhang {\it et al}. \cite{Alex}, with artificial one electron doping, they have shown that the Fermi level can be shifted to conduction band gap in BaBiO$_3$. For practical purposes, we feel that fluorine doping can be of some help due to the following reasons: (I) It has similar covalent radius (0.66 {\AA} as against 0.64 {\AA} of O). There are many examples to site from the literature concerning F doping at O site. Few of them are LaO$_{1-x}$F$_{x}$FeAs (x = 0.5), CeO$_{1-x}$F$_{x}$BiS$_2$ (x = 0-0.6)\cite{Lao,Ceo}. (II) More importantly F dopant does not change the  orbital symmetry since the same 2s and 2p states are involved in the band formation. Hence, it is expected that the band topology will not be affected due to F doping.  \\

\textbf{Acknowledgments:}
The authors acknowledge the computational
resources provided by HPCE, IIT Madras. This work is supported by
Department of Science and Technology, India through Grant
No. EMR/2016/003791.

\begin{widetext}

\section*{APPENDIX-I: Tight-Binding Model}
 The basis set for the TB Hamiltonian in Eq. (1)  consists of  the one Bi-s, three Bi-p,  and the nine O-p (three per each oxygen anion) orbitals. The rest are neglected as they neither lie in the vicinity of the Fermi level nor participate in the formation of the topological invariant states. Also the hopping interactions are confined to the nearest neighbor and next nearest neighbor coordination except the case of O. This is due to the fact that O-p - O-p interaction is not significant in forming the highly dispersive states at $\Gamma$ in the valence band and at R at the conduction band. Therefore, TB Hamiltonian matrix, constructed with the help of Slater-Koster hopping integrals (SKI) \cite{Koster}, includes Bi-s - O-p and Bi - Bi interactions. The relevant SKIs are reproduced here.

\begin{eqnarray}
E_{s,x} &=& lt_{sp\sigma} \nonumber \\
E_{x,x} &=& l^2t_{pp\sigma}-(1-l^2)t_{pp\pi} \nonumber \\
E_{x,y} &=& lmt_{pp\sigma}-lmt_{pp\pi} \\
E_{x,z} &=& lnt_{pp\sigma}-lnt_{pp\pi},\nonumber
\end{eqnarray}

where, $x,y$, and $z$ represent $p_x,p_y$, and $p_z$ orbitals, respectively, and $l,m,n$ are direction cosines. The parameter $t$ quantifies the covalent interactions of various types ($\sigma$ and $\pi$). The spin independent TB Hamiltonian matrix, with the basis set in the order $\{|s^{Bi}\rangle$ $|p^{Bi}_{x}\rangle$, $|p^{Bi}_{y}\rangle$, $|p^{Bi}_{z}\rangle$  $|p^{O1}_{x}\rangle$, $|p^{O1}_{y}\rangle$, $|p^{O1}_{z}\rangle$, $|p^{O2}_{x}\rangle$, $|p^{O2}_{y}\rangle$, $|p^{O2}_{z}\rangle$, $|p^{O3}_{x}\rangle$, $|p^{O3}_{y}\rangle$, $|p^{O3}_{z}\rangle$$\}$, takes the form:

\begin{equation}
H = \left( \begin{array}{cc}
M_{4\times4}^{Bi-Bi}&M_{4\times9}^{Bi-O}\\\\
M_{4\times9}^{Bi-O \dag}&M_{9\times9}^{O-O}
\end{array}  \right)
\end{equation}

The individual blocks, of the matrix H, are as follows,
\begin{equation}
M_{4\times4}^{Bi-Bi} = \left( \begin{array}{cccc}
\epsilon_s+f_{1}	& 2it_{sp\sigma}^{Bi-Bi}\sin(k_xa)& 2it_{sp\sigma}^{Bi-Bi}\sin(k_ya) &2it_{sp\sigma}^{Bi-Bi}\sin(k_za)  \\
-2it_{sp\sigma}^{Bi-Bi}\sin(k_xa)	&\epsilon_{p1}+f_2 &	0	&0	\\
-2it_{sp\sigma}^{Bi-Bi}\sin(k_ya) &0&\epsilon_{p1}+f_3&0\\
-2it_{sp\sigma}^{Bi-Bi}\sin(k_za)	&0	&0	&\epsilon_{p1}+f_4 \\
\end{array}  \right)
\end{equation}

\begin{equation}
M_{4\times9}^{Bi-O} = \left( \begin{array}{ccccccccc}
t_{sp\sigma}^{Bi-O}S_x 	&0	&0	&0	& t_{sp\sigma}^{Bi-O}S_y	&0	&0	&0	& t_{sp\sigma}^{Bi-O}S_z \\
t_{pp\sigma}^{Bi-O}C_x	&0	&0	&t_{pp\pi}^{Bi-O}C_y   &	0	&0	&t_{pp\pi}^{Bi-O}C_z&	0	&0\\
0&t_{pp\pi}^{Bi-O}C_x&0	&0&t_{pp\sigma}^{Bi-O}C_y&0 &0&t_{pp\pi}^{Bi-O}C_z &0\\
0	&0	&t_{pp\pi}^{Bi-O}C_x	&	0&0&t_{pp\pi}^{Bi-O}C_y 	&0	&0	&t_{pp\sigma}^{Bi-O}C_z\\
\end{array}  \right)
\end{equation}

\begin{equation}
M_{9\times9}^{O-O} = \left( \begin{array}{ccccccccc}

\epsilon_{p2}&0&0&0&0&0&0&0&0\\
0&\epsilon_{p2}&0&0&0&0&0&0&0\\
0&0&\epsilon_{p2}&0&0&0&0&0&0\\
0&0&0&\epsilon_{p2}&0&0&0&0&0\\
0&0&0&0&\epsilon_{p2}&0&0&0&0\\
0&0&0&0&0&\epsilon_{p2}&0&0&0\\
0&0&0&0&0&0&\epsilon_{p2}&0&0\\
0&0&0&0&0&0&0&\epsilon_{p2}&0\\
0&0&0&0&0&0&0&0&\epsilon_{p2}

\end{array}  \right)
\end{equation}

Here $\epsilon_s$, $\epsilon_{p1}$, and $\epsilon_{p2}$ are on-site energies of Bi-s, Bi-p, and O-p orbitals respectively. The terms $C_x$ and $S_x$ are short notation for $2\cos(k_xa/2)$ and $2i\sin(k_xa/2)$  and $f_i$ $(i=1,2,3,4)$ arises from Bi-Bi second neighbor interactions,  are given by
\begin{eqnarray}
f_1 &=& 2t_{ss}^{Bi-Bi}(cos(k_xa)+cos(k_ya)+cos(k_za)) \nonumber \\
f_2 &=& 2t_{pp\sigma}^{Bi-Bi}cos(k_xa)+2t_{pp\pi}^{Bi-Bi}[cos(k_ya)+cos(k_za)] \nonumber \\
f_3 &=& 2t_{pp\sigma}^{Bi-Bi}cos(k_ya)+2t_{pp\pi}^{Bi-Bi}[cos(k_xa)+cos(k_za)]\\
f_4 &=& 2t_{pp\sigma}^{Bi-Bi}cos(k_za)+2t_{pp\pi}^{Bi-Bi}[cos(k_xa)+cos(k_ya)] \nonumber
\end{eqnarray}

 The {\it k}-dependency arises through the Bloch summation:
 
 \begin{equation}
    h_{\alpha \beta} (k) = \sum_{l\neq j} t_{j\alpha l\beta} e^{i\vec{k} \cdot (\vec{R}_l - \vec{R}_j)},
 \end{equation}
 
 where $j$ and $\alpha$ respectively represent site and orbital indices and $R_j$ is the position vector for the  $j$-th site. The terms $t_{j\alpha l\beta}$ are constructed from Eq. (3). 

 The SOC component of the Hamiltonian, acting on the Bi-p states, is expressed as:
 
\begin{align}
\begin{split}
H_{SOC}& =\lambda{\textbf{L}}\cdot{\textbf{S}}
\end{split}
\end{align}

\begin{align}
\begin{split}
\langle L\cdot S\rangle& =\frac{1}{2}\langle J^2 - L^2 -S^2 \rangle\\ &= \frac{\hbar^2}{2}(j(j+1)-l(l+1)-s(s+1))
\end{split}
\end{align}

To operate the SOC Hamiltonian on the p-orbitals, we need to express them as a linear combination of total angular momentum states ($\Phi_{j,m_j}$).

\begin{eqnarray}
p_x\uparrow &= &\frac{1}{\sqrt{2}}\Big(\Phi_{\frac{3}{2},\frac{3}{2}}+\frac{1}{\sqrt{3}}\Phi_{\frac{3}{2},-\frac{1}{2}} - \sqrt{\frac{2}{3}}\Phi_{\frac{1}{2},-\frac{1}{2}}\Big) \nonumber \\
p_y\uparrow &= &-\frac{i}{\sqrt{2}}\Big(\Phi_{\frac{3}{2},\frac{3}{2}}-\frac{1}{\sqrt{3}}\Phi_{\frac{3}{2},-\frac{1}{2}} + \sqrt{\frac{2}{3}}\Phi_{\frac{1}{2},-\frac{1}{2}}\Big) \nonumber \\
p_z\uparrow &= &\sqrt{\frac{2}{3}}\Phi_{\frac{3}{2},-\frac{1}{2}} - \sqrt{\frac{1}{3}}\Phi_{\frac{1}{2},-\frac{1}{2}} \nonumber \\
p_x\downarrow &= &\frac{1}{\sqrt{2}}\Big(\frac{1}{\sqrt{3}}\Phi_{\frac{3}{2},\frac{1}{2}} + \sqrt{\frac{2}{3}}\Phi_{\frac{1}{2},\frac{1}{2}} + \Phi_{\frac{3}{2},-\frac{3}{2}}\Big)\\
p_y\downarrow &= &-\frac{i}{\sqrt{2}}\Big(\frac{1}{\sqrt{3}}\Phi_{\frac{3}{2},\frac{1}{2}} + \sqrt{\frac{2}{3}}\Phi_{\frac{1}{2},\frac{1}{2}} - \Phi_{\frac{3}{2},-\frac{3}{2}}\Big) \nonumber \\
p_z\downarrow &= &\sqrt{\frac{2}{3}}\Phi_{\frac{3}{2},-\frac{1}{2}} + \sqrt{\frac{1}{3}}\Phi_{\frac{1}{2},-\frac{1}{2}} \nonumber
\end{eqnarray}

With the aid of Eqs. (10) and (11), we obtain the matrix elements for the SOC Hamiltonian with the basis set in the order: $|p_{x}^{Bi}\uparrow\rangle$, $|p^{Bi}_y\uparrow\rangle$, $|p^{Bi}_z\uparrow\rangle $, $|p^{Bi}_x\downarrow\rangle $, $|p^{Bi}_y\downarrow\rangle $, $|p^{Bi}_z\downarrow\rangle $

\begin{equation}
H_{SOC}
 =  \lambda\left( \begin{array}{cccccc}
0&-i&0&0&0&1 \\
i&0&0&0&0&-i\\
0&0&0&1&-i&0\\
0&0&1&0&i&0\\
0&0&i&-i&0&0\\
1&i&0&0&0&0
\end{array}  \right)
\end{equation}

 The exact diagonalization of the Hamiltonian gives the band dispersion which is further fitted  with that of DFT with RMS deviation tolerance less than 0.25 to yield the optimized tight binding parameters. These parameters  in the absence of SOC are listed in Table I and the same in the presence of SOC are listed in Table II. Table II also lists the SOC strength $\lambda$. As expected, with inclusion of SOC, the only second neighbor Bi-Bi interaction parameters changes with respect to that in Table I.  
 
\begin{table*}
\centering
\caption{The optimized tight-binding parameters in the absence of SOC.}
\begin{ruledtabular}
\begin{tabular}{ p{1.2cm}p{1.5cm}p{1cm}p{1cm} p{1cm}p{1cm}p{1cm}p{1cm}p{1cm}p{1cm}p{1.2cm}p{1cm} }
charge & A & $E_{Bi-s}$ &$E_{Bi-p}$ &$E_{O-p}$&$t_{sp}$ &$t_{pp\sigma}$ & $t_{pp\pi}$ &$t_{ss}^{Bi-Bi}$ & $t_{sp\sigma}^{Bi-Bi}$&$t_{pp\sigma}^{Bi-Bi}$&$t_{pp\pi}^{Bi-Bi}$\\
 \hline\\
+1&Na	&-4.78	&5.32	&-1.55	 &2.08&	2.75&	-0.7&-0.04&0.06&-0.12&0.08\\
&K     &-4.4   &5.7    &-1.6   &2.2     &2.95&	-0.75&-0.04&0.02&-0.08&0.16\\
&Rb    &-4.6	  &5.5	&-1.5	&2.1 &	2.8 &	-0.7&-0.03&0.07&-0.01&0.16\\
&Cs& -5&	5.1&	-1	&2.1 &	2.8&-0.75&-0.07&0.11&-0.14&0.2\\\\	
+2& Mg& -6.6 &	3.5&	-2.65	 &2.1&	2.75&	-0.85&-0.02&-0.0&-0.05&0.04\\
& Ca&-6.2&	3.9&	-2.5	&2.1&	2.8&	-1.4&-0.04&-0.04&-0.1&0.55\\
& Sr & -6.3 &	3.8	&-2.2&	2&	2.7&	-1.35&-0.046&-0.08&-0.22&0.45\\
& Ba &-6.3	 & 3.8&	-2.2&	2&	2.82&	-1.35&-0.04&-0.1&-0.25&0.5\\

\end{tabular}
\end{ruledtabular}
\end{table*}

\begin{table*}
\centering
\caption{The optimized tight-binding parameters in the presence of SOC.}
\begin{ruledtabular}
\begin{tabular}{ p{1cm}p{1.3cm}p{1cm}p{1cm} p{1cm}p{1cm}p{1cm}p{1cm}p{1cm}p{1cm}p{1cm}p{1cm}p{0.8cm} }

charge & A & $E_{Bi-s}$ &$E_{Bi-p}$ &$E_{O-p}$&$t_{sp}$ &$t_{pp\sigma}$ & $t_{pp\pi}$& $t_{ss}^{Bi-Bi}$ & $t_{sp\sigma}^{Bi-Bi}$&$t_{pp\sigma}^{Bi-Bi}$&$t_{pp\pi}^{Bi-Bi}$&$\lambda$ \\
 \hline\\
+1&Na& -4.78	&5.32&	-1.55&	2.08&	2.61&	-0.9	&-0.05& 0.05&-0.05&0.1&0.7 \\
& K & -4.4&	5.7&	-1.6	&2.15&	2.82&	-0.7&-0.05& 0.05&-0.05&0.15&	0.7\\
& Rb & -5.2&	4.9&	-0.5	&2.08&	2.9&	-1.15&0.07&0.05&-0.05&0.1&	0.6\\
& Cs & -4.9&	5.2&	-0.8&	2&	3&	-0.85&-0.02&-0.12&-0.15&0.25&	0.7\\\\

+2& Mg &-6.4	&3.7&	-2.3&	2.1&	2.75&	-1.3&0.02&0.05&-0.4&0.3&	0.7\\
& Ca &-6.4&	3.7& 	-2.3&	2&	2.75&	-1.5&-0.03&0.05&-0.3&0.5&	0.5\\
&Sr&-6.4	&3.7&	-2.3	&2.0	&2.7&	-1.5&-0.03&0.05&-0.3&0.5&	0.45\\
&Ba&-6.5&	3.6&	-2.3&	2&	2.7&	-1.3&0.03&0.05&0.3&0.2&	0.45\\
\end{tabular}
\end{ruledtabular}
\end{table*}

\section*{APPENDIX-II: Electronic structure of $\mathbf{{\it A}BiO_3}$}

\begin{figure*}[h]
\centering
\includegraphics[angle=-0.0,origin=c,height=10cm,width=16.0cm]{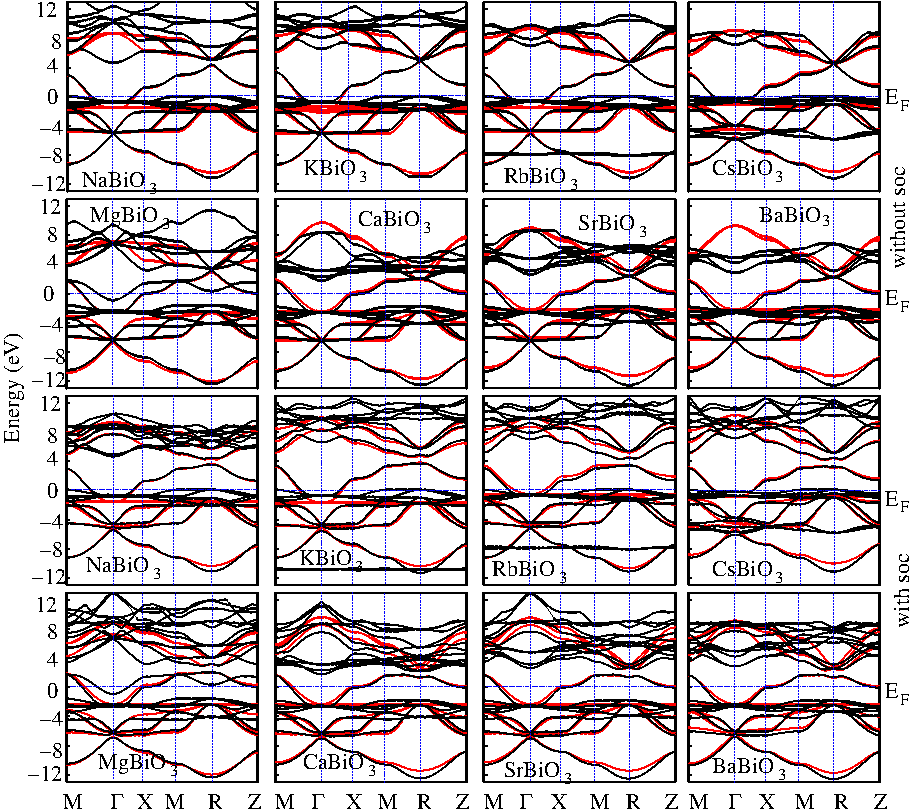}
\caption{Bulk band structure of {\it A}BiO$_3$  without SOC (upper two rows) and with SOC (lower two rows). Black and red lines represent bands obtained from DFT and TB,  respectively. TB bands are composed of Bi-$\{$s,p $\}$ and O-$\{$p$\}$ orbitals. There is an excellent agreement between these two sets of bands. As discussed in the main text, narrow (or vanishing) band gaps at $\Gamma$ in the valence band and at R in the conduction band suggest the possibility of formation of two topologically invariant surface states. The SOC either creates a gap at $\Gamma$ and R or enhances the already existing gaps at these high symmetry points.}
\label{fig:7}
\end{figure*}
\begin{figure*}[h]
\centering
\includegraphics[angle=-0.0,origin=c,height=5.5cm,width=16.0cm]{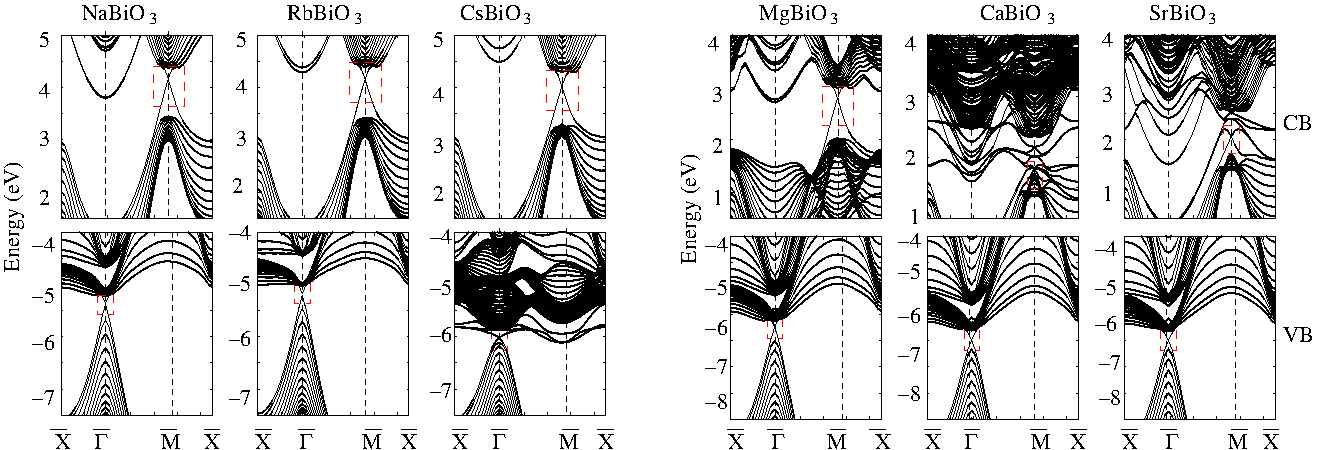}
\caption{DFT+SOC surface band structure  for {\it A}BiO$_3$ (A=Na, Rb, Cs, Mg, Ca and Sr) along the high symmetry points of the surface Brillouin zone. The first row represents the conduction bands and the  second row represents the valence bands.  Except for {\it A} = Na the surface electronic structure is obtained for a 15 unit cell thick {\it A}BiO$_3$. For Na 13 unit cell is used.   The Dirac type surface states, highlighted with dashed rectangles,  appear  at two time reversal invariant momenta points $\bar{\Gamma}$ and $\bar{M}$.}
\label{fig:8}
\end{figure*}

To avoid repetition, the bulk and surface electronic structure of each member of the ABiO$_3$ family are not shown in the main text. However, for reference they are shown in this appendix. 
The captions of the figures and the tables are complete and self explanatory. The orbital weights, as obtained from the TB calculations, are listed in Table III (without SOC)  and Table IV (with SOC). With inclusion of SOC the s-p band inversion occurs at $\Gamma$ for all the compounds. However, at $R$, in the conduction band, the s-p band inversion is already seen without SOC for the compounds CaBiO$_3$, SrBiO$_3$ and BaBiO$_3$. For the rest of the compounds SOC creates the band inversion. 

\begin{table*}
\centering
\caption{Orbital weight factors of different ABiO$_3$ compounds without spin orbit coupling.}
\begin{ruledtabular}
\begin{tabular}{ccccccccc} 

 & \multicolumn{4}{c}{CB orbital weight factors\%}  & \multicolumn{4}{c}{VB orbital weight factor \%} \\\cline{2-5}\cline{6-9}
    & Band index  & Bi-s & Bi-p & O-p &Band index & Bi-s & Bi-p & O-p\\
    \hline\\
\raisebox{-1.5ex}[0pt]{Na}	& 1	&40	&0	&60	& 1	&100	&0 	&0 \\
	                       & 2	&0	&100	&0	& 2	&0	& 25	& 75\\ 
\raisebox{-1.5ex}[0pt]{K}	& 1	&41.7	&0	&58.3	& 1	&0	&26.3 	&76.3 \\
	                       & 2	&0	&100	&0	& 2	&0	&26.3 	&76.3 \\
\raisebox{-1.5ex}[0pt]{Rb}	& 1	&40	&0	&60	& 1	&100	&0 	&0 \\ 
 	                       & 2	&0	&100	&0	& 2	&0	&23 	&77 \\
 	                       
\raisebox{-1.5ex}[0pt]{Cs}	& 1	&0	&100	&0	& 1	&100	& 0	&0 \\
	                       & 2	&0	&100	&0	& 2	&0	&25.8 	&74.2 \\
\raisebox{-1.5ex}[0pt]{Mg}	& 1	&37	&0	&63	& 1	&100	&0 	&0 \\ 
 	                       & 2	&0	&	100&0	& 2	&	0& 27	&73 \\
 	                      
\raisebox{-1.5ex}[0pt]{Ca}	& 1	&0	&100	&0	& 1	&100	&0 	&0 \\
	                       & 4	&39.8	&	0&60.2	& 2	&0	&24.1 	&75.9 \\
\raisebox{-1.5ex}[0pt]{Sr}	& 1	&0	&100	&0	& 1	&100	&0 	&0 \\ 
 	                       & 4	&36.7	&0	&63.3	& 2	&0	& 25.6	&74.4 \\	                       
\raisebox{-1.5ex}[0pt]{Ba}	& 1	&0	&100	&0	& 1	&100	&0 	&0 \\ 
 	                       & 4	&36.6	&0	&63.4	& 2	&0	& 25.9	&74.1 \\
\end{tabular}
\end{ruledtabular}
\end{table*}

\begin{table*}
\centering
\caption{Orbital weight factors of different ABiO$_3$ compounds with spin orbit coupling.}
\begin{ruledtabular}
\begin{tabular}{ccccccccc} 

 & \multicolumn{4}{c}{CB orbital weight factors\%}  & \multicolumn{4}{c}{VB orbital weight factor \%} \\\cline{2-5}\cline{6-9}
    & Band index  & Bi-s & Bi-p & O-p &Band index & Bi-s & Bi-p & O-p\\
    \hline\\
\raisebox{-1.5ex}[0pt]{Na}	& 1	& 0	&	100 & 0	& 1	&	0&27 	&73 \\
	                       & 2	&40	&0	&60	& 2	&100	& 0	&0 \\
\raisebox{-1.5ex}[0pt]{K}	& 1	&0	&100	&0	& 1	&0	&26.5 	&73.5 \\
	                       & 2	&41.7	&0	&58.3	& 2	&100	&0 	&0 \\ 
\raisebox{-1.5ex}[0pt]{Rb}	& 1	&0	&	100&0	& 1	&0	&34 	& 66\\ 
 	                       & 2	&33.3	&0	&66.7	& 2	&100	&0 	&0 \\
\raisebox{-1.5ex}[0pt]{Cs}	& 1	&0	&100	&0	& 1	&0	&31 	&69 \\
	                       & 2	&36.2	&0	&63.8	& 2	&100	&0 	&0 \\
	                       
\raisebox{-1.5ex}[0pt]{Mg}	& 1	&0	&100	&0	& 1	&0	&32.3 	&67.7 \\ 
 	                       & 2	&36	&0	&64	& 2	&100	&0 	&0 \\
\raisebox{-1.5ex}[0pt]{Ca}	& 1	&0	&100	&0	& 1	&0	&29 	&71 \\
	                       & 4	&36.4	&0	&63.6	& 2	&100	&0 	&0 \\
\raisebox{-1.5ex}[0pt]{Sr}	& 1	&0	&	100&0	& 1	&0	&28.6 	&71.4 \\ 
                            & 4	&36.4	&0	&63.6	& 2	&100	&0 	&0 \\                      
\raisebox{-1.5ex}[0pt]{Ba}	& 1	&0	&100	&0	& 1	&0	&28 	&72 \\ 
 	                       & 4	&35	&0	&65	& 2	&100	&0 	&0 \\
\end{tabular}
\end{ruledtabular}
\end{table*}

\section*{APPENDIX-III: Robustness of surface TI states of $\mathbf{{\it A}BiO_3}$ against Exchange Correlation approximation and Surface deformation.}

While DFT calculations are very promising in examining the topological order of materials, a study carried out by Vidal {\it et al}.\cite{Vidal}  shows that the band-gap underestimation, due to exchange-correlation approximation may yield false topological order. For example, DFT+GGA calculations suggests compounds like LuPTSb, YPtSb are topological insulators\cite{Sawai}. However, with the incorporation of more accurate quasiparticle GW approximation, there is an increase in the bulk band gap and these compounds are found to be normal insulator\cite{Vidal}. Also the robustness of TI states should be checked against surface deformation

The effect of exchange-correlation approximation is investigated on two prototype compounds KBiO$_3$ and BaBiO$_3$. While K has +1 charge state, Ba has +2 charge state. Instead of GW approximation, we have applied the less expensive modified Becke-Johnson potential (mBJ) as a correction to GGA. Tran and Blaha\cite{Tran} have shown that mBJ corrections provide the same order of band gap as by hybrid functionals and GW methods for all types of solids(wide band gap insulators, sp- semiconductors, and strongly correlated 3d transition-metal oxides). The band structures and the weights of the relevant orbitals for these two compounds obtained with GGA + mBJ + SOC  are shown in Fig.~\ref{fig:9}. The Bi s-p band inversion, both in conduction and valence band  is not affected with the inclusion of mBJ. In fact the surface band structure of 11 unit-cell thick slabs of KBiO$_3$ and BaBiO$_3$ , shown in Fig.~\ref{fig:10} reconfirm the existence of conduction and valence TI Dirac states. Furthermore, the variation of surface band gap as a function of slab thickness, shown in Fig.~\ref{fig:11}, indicates that the surface penetration length, above which the interaction between the top and bottom surface states ceases, remains the same both for GGA and GGA+mBJ approximations.

\begin{figure}[h]
\includegraphics[angle=-0.0,origin=c,height=6cm,width=16cm]{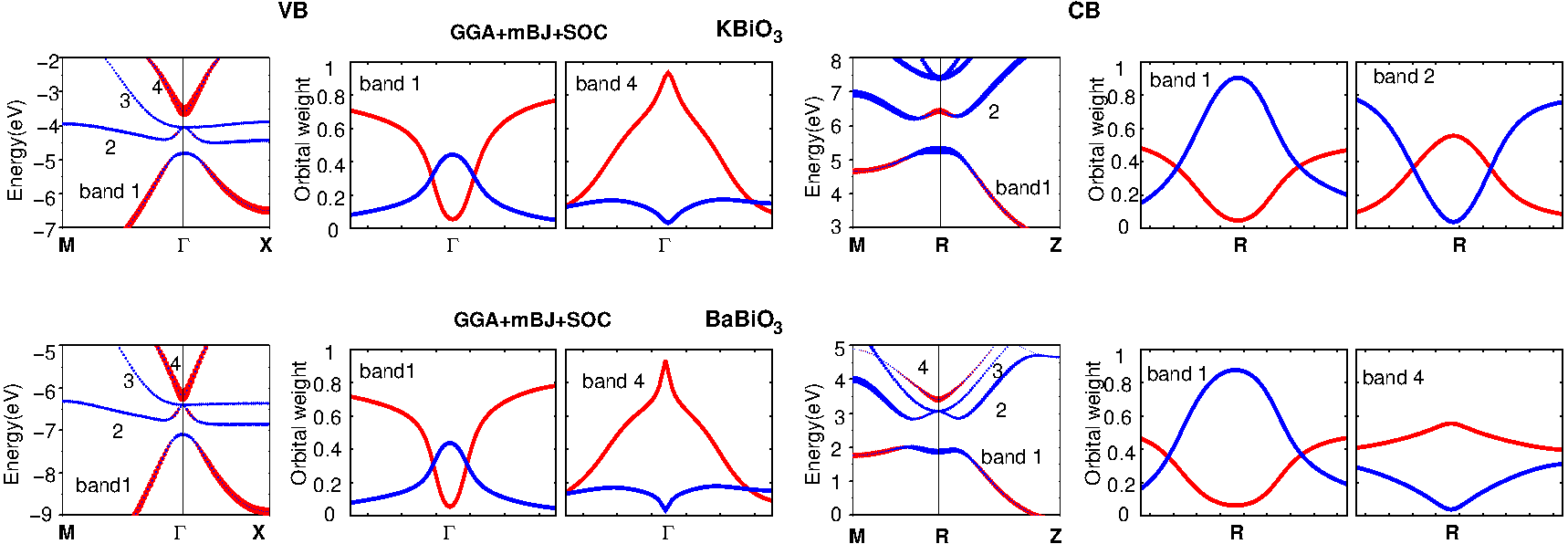}
\caption{First and fourth columns: GGA + mBJ + SOC Band structures of KBiO$_3$ and BaBiO$_3$ in the vicinity of bulk valence band (VB) gap (around $\Gamma$) and bulk conduction band (CB) gap (around $R$). The red and blue identify the Bi-s and Bi-p characters. Second, third, fifth and sixth columns: The Bi-s and -p orbital weights of the indicated bands are shown to quantify the band inversion. Even though O-p characters are present throughout, their contributions are not shown here as they have no role in the band inversion. It may be noted that the mBJ correction can shift the bands participating in the inversion. For example, in VB with mBJ the band inversion occurs between band-1 and band-4 where as without mBJ correction the inversion occurs between band-1 and band-2 (see Fig. 4). However, the formation of surface Dirac states is not affected by such shifting.} 
\label{fig:9}
\end{figure}

\begin{figure}[h]
\includegraphics[angle=-0.0,origin=c,height=8cm,width=14cm]{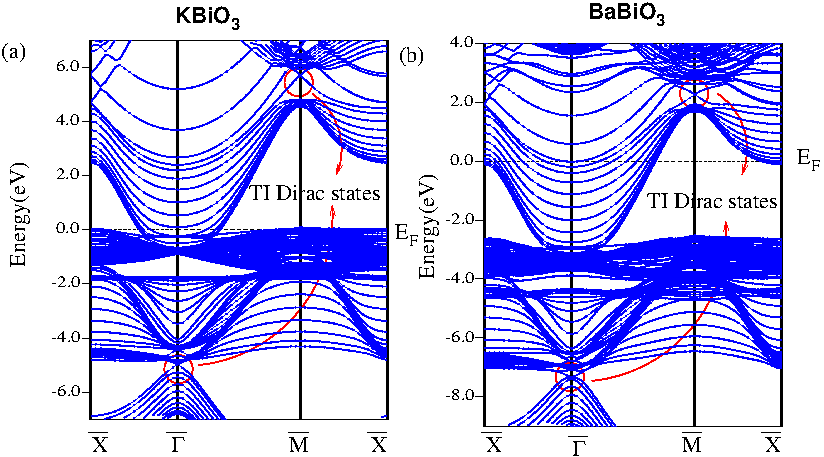}
\caption{Surface band structure of 11 unit slabs of (a) KBiO$_3$ and (b)  BaBiO$_3$. calculated using GGA + mBJ +SOC formalism.}
\label{fig:10}
\end{figure}

\begin{figure}[h]
\includegraphics[angle=-0.0,origin=c,height=8cm,width=10cm]{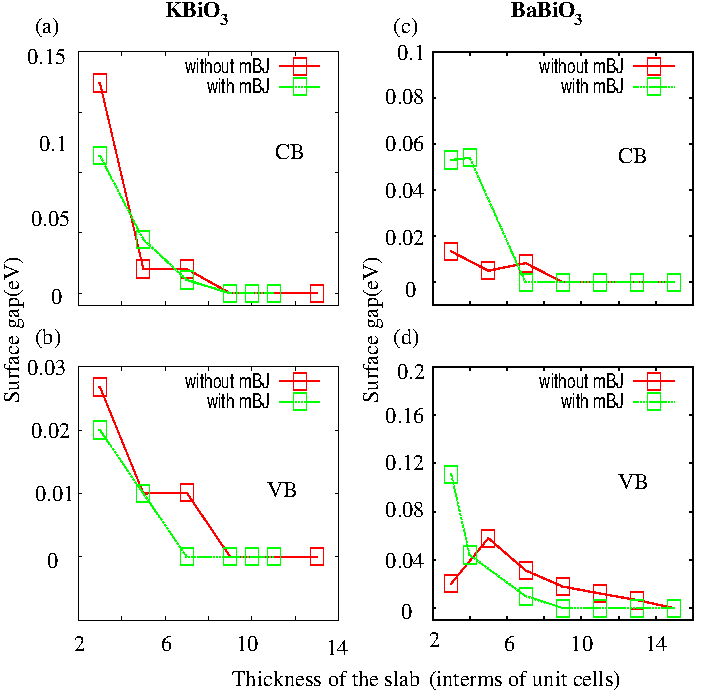}
\caption{Variation of surface Dirac gap of KBiO$_3$ and BaBiO$_3$. The CB Dirac gaps are shown in (a) and (c) and the VB Dirac gaps are shown in (b) and (d). } 
\label{fig:11}
\end{figure}

\subsection{ Effect of deformation on the TI behavior in {\it A}BiO$_3$}

It is known that the many members of the family of ABiO$_3$ stabilizes in a distorted cubic structure. A good example is BaBiO$_3$. It is experimentally found that this compound has a breathing and tilting  mode of distortions involving the BiO$_6$ octahedra. However, the tilting mode can be suppressed when grown as thin films on an appropriate substrate (e.g. MgO \cite{Kei}). Therefore, it is significant to examine the nature of s-p band inversion under the breathing mode of distortion.

We considered the breathing mode of distortion of the BiO$_6$ octahedra in a G-type pattern in BaBiO$_3$ as shown in Fig. \ref{fig12}(a). The total energy as a function of O anion displacement ($\Delta Q$) is estimated and we find that there exists a double minima curve [see Fig. \ref{fig12}(b)] which is in agreement with earlier reports \cite{Rabe}. The system stabilizes with $\Delta Q \approx \pm 0.04$\AA. The corresponding spin-orbit coupled band structure is shown in Fig. \ref{fig12}(c). Like the undistorted structure, it exhibits bulk band gap as well as s-p band inversion at 2 eV above and -7 eV below E$_F$. 

\begin{figure}[h]
\begin{center}
\hspace{0.6cm}
\includegraphics[angle=-0.0,origin=c,height=7cm,width=11cm]{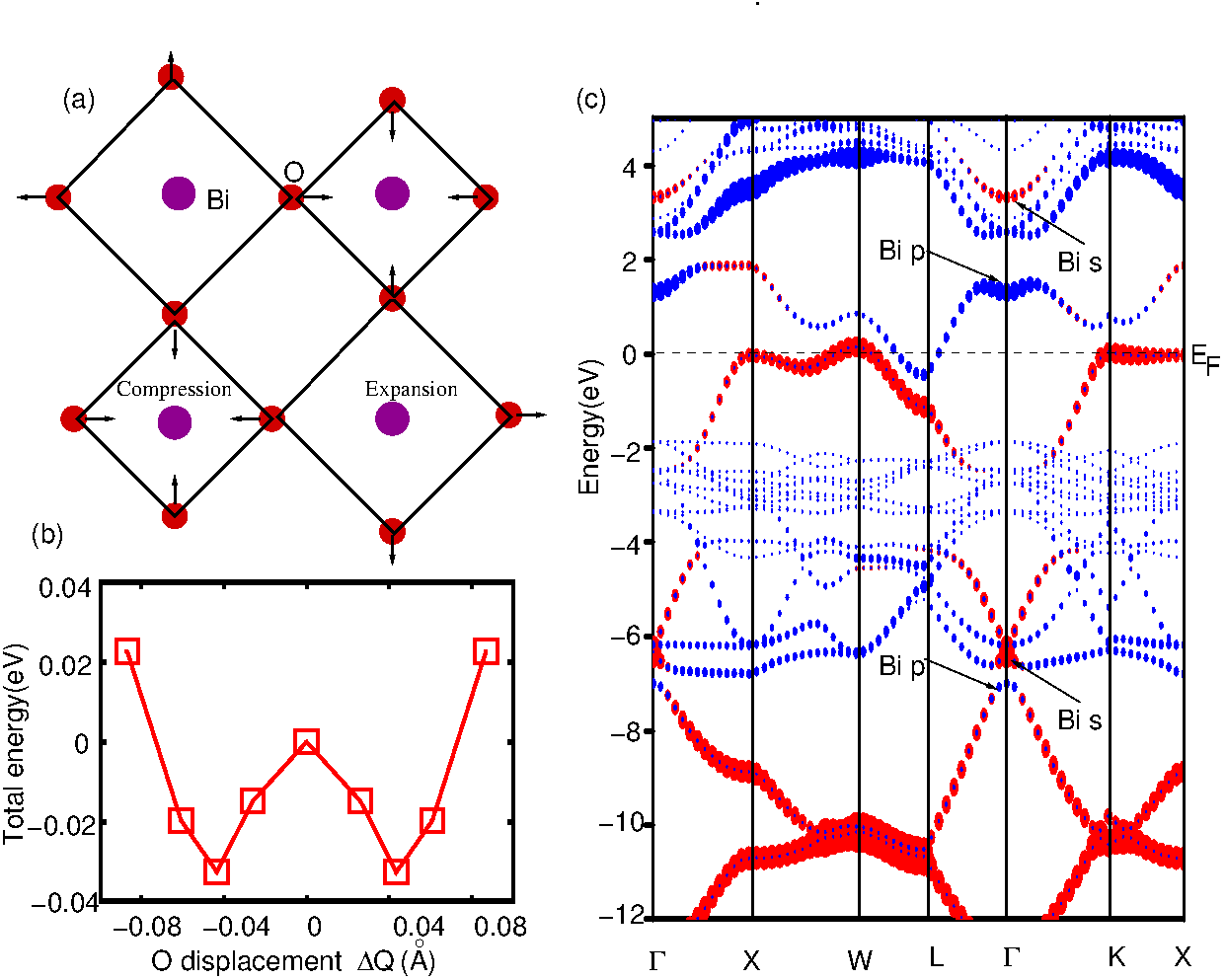}
\caption{(a) Schematic illustration of breathing mode distortion of the BiO$_6$ octahedra in BaBiO$_3$. (b) Variation of total energy as a function of breathing distortion strength. (c)The orbital projected spin-orbit coupled band structure corresponding to minimum breathing distortion. The contribution from Bi-s and Bi-p orbitals are represented through red and blue circles, respectively. Like the cubic perovskite, the distorted structure shows s-p band inversion. The band structure is calculated using a two formula unit fcc primitive cell incorporating the breathing mode of distortion. It may be noted that the high symmetry point R of the cubic Brillouin zone maps to $\Gamma$ of the new fcc Brillouin zone.  } 
\label{fig12}
\end{center}
\end{figure}

The band structure of the distorted BaBiO$_3$ reaffirms the robustness of s-p band inversion and therefore possible formation of TI surface Dirac states. To examine the surface deformation of this distorted BaBiO$_3$, we doubled the cell along $x$ and $y$ and constructed a 21 layer thick slab consisting of 204 atoms. As the FP-LAPW method is computationally expensive for such a large system, the calculations (structural relaxation and band structure) are carried out using pseudopotential approximation as implemented in VASP\cite{Kresse,Joubert}.

\begin{figure}[H]
\begin{center}
\hspace{-0.6cm}
\includegraphics[angle=-0.0,origin=c,height=7.5cm,width=13cm]{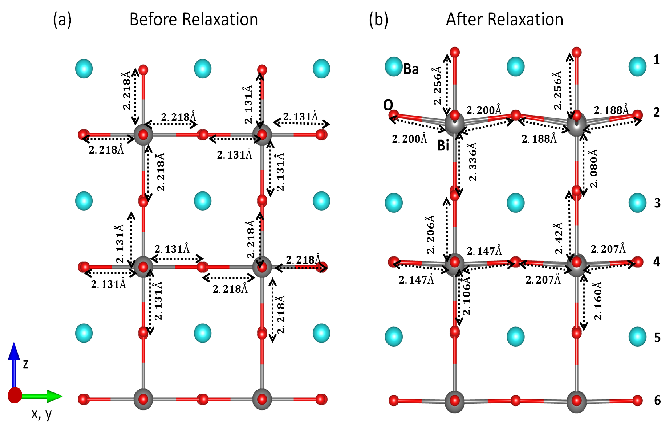}
\caption{(a) Top six layers of the unrelaxed slab constructed out of the bulk having breathing mode of distortion with $\Delta Q = \pm 0.04$ \AA (see Fig. \ref{fig12}). (b) The corresponding relaxed structure. The change in the bond lengths demonstrates the surface deformation. The atoms lying six layers away from the surface experience negligible displacement.}
\label{fig13}
\end{center}
\end{figure}

\begin{figure}[h]
\begin{center}
\hspace{0.2cm}
\includegraphics[angle=-0.0,origin=c,height=6cm,width=14cm]{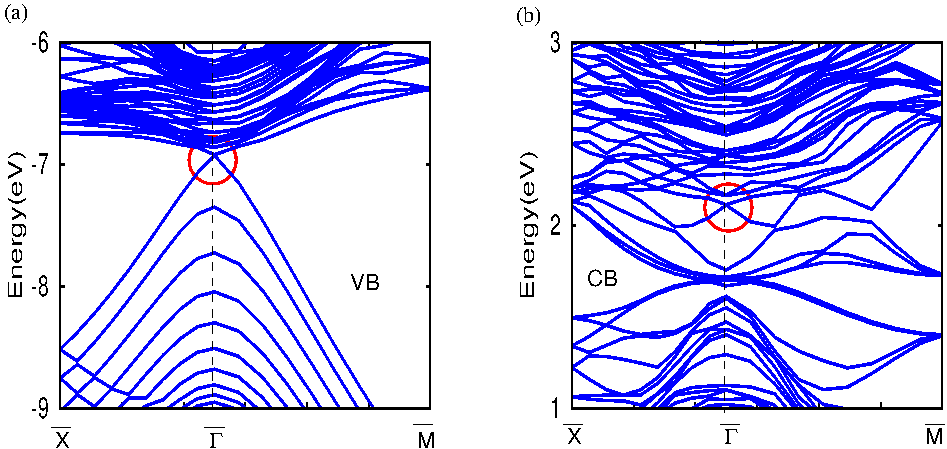}
\caption{Surface band dispersions are obtained after the relaxation of 21 layers thick slab with breathing distortion. The band structure with spin orbit coupling in the valence band(VB) spectrum (a) and conduction band(CB) spectrum (b) are drawn along the high symmetry path of the surface Brillouin zone. The linearly dispersed  Dirac bands are encircled.} 
\label{fig14}
\end{center}
\end{figure}

The starting structure with equilibrium breath-in and breath-out distortion is shown in Fig. \ref{fig13}(a) and the corresponding relaxed structure is shown in Fig. \ref{fig13}(b). Only the first six layers from the surface are shown, as distortion in the further inner layers are negligible. We find that except O, the other two heavy ions have negligible displacements after relaxation. Therefore, the strength of distortion is presented through the Bi-O bond length. The distortions are more along the $z-axis$ compared to the $xy$ plane. Also it is observed that maximum distortion occurs in the first three layers. However, the nature of the  breathing mode remains unaltered. The SOC driven surface band structure for the relaxed structure is shown in Fig. \ref{fig14}. Since the s-p band inversion and hence perspective TI states occur at $\sim$ -7 eV and $\sim$ 2 eV with respect to the Fermi level, in Fig. \ref{fig14} we have shown the band structures in these vicinities only. As in the case of undistorted  cubic surface,  the valence band structure shows the linear band crossing at around -7 eV and the conduction band structure shows linear band crossing at 2 eV confirming the presence of both V-TI and C-TI states. This implies that the formation of surface TI states are invariant under adiabatic surface deformation for BaBiO$_3$. It is expected that the other members of the {\it A}BiO$_3$ family will behave similarly. 

\end{widetext}

\clearpage
\bibliography{paper}

\end{document}